\documentclass[pra,twocolumn,10pt,notitlepage,superscriptaddress,tightenlines,amsmath,amssymb,showpacs,eqsecnum]{revtex4-1}

\usepackage{bm,graphicx,hyperref,upgreek,empheq}

\def\UpsilonT{\Upsilon_{\!T}}

\begin{document}

\title{Entanglement-based perturbation theory\\ for highly anisotropic Bose-Einstein condensates}

\author{Alexandre B.~Tacla}
\affiliation{
Center for Quantum Information and Control, MSC07--4220, University of New Mexico,
Albuquerque, New Mexico 87131-0001, USA}

\author{Carlton M.~Caves}
\affiliation{
Center for Quantum Information and Control, MSC07--4220, University of New Mexico,
Albuquerque, New Mexico 87131-0001, USA}
\affiliation{Centre for Engineered Quantum Systems, School of Mathematics and Physics, University of Queensland, Brisbane, Queensland 4072, Australia}

\begin{abstract}
	
We investigate the emergence of three-dimensional behavior in a reduced-dimension Bose-Einstein condensate trapped by a highly anisotropic potential.  We handle the problem analytically by performing a perturbative Schmidt decomposition of the condensate wave function between the tightly confined (transverse) direction(s) and the loosely confined (longitudinal) direction(s).  The perturbation theory is valid when the nonlinear scattering energy is small compared to the transverse energy scales.  Our approach provides a straightforward way, first, to derive corrections to the transverse and longitudinal wave functions of the reduced-dimension approximation and, second, to calculate the amount of entanglement that arises between the transverse and longitudinal spatial directions.  Numerical integration of the three-dimensional Gross-Pitaevskii equation for different cigar-shaped potentials and experimentally accessible parameters reveals good agreement with our analytical model even for relatively high nonlinearities.  In particular, we show that even for such stronger nonlinearities the entanglement remains remarkably small, which allows the condensate to be well described by a product wave function that corresponds to a single Schmidt term.

\end{abstract}

\maketitle

\section{Introduction}

Three-dimensional Bose-Einstein condensates (BECs) confined in highly anisotropic traps are known to exhibit lower-dimensional behavior when the number of condensed atoms is well below a critical value~\cite{gorlitz01}. Under such conditions, the tightly confined dimension(s) can be effectively neglected because the characteristic transverse energy scale far exceeds the scattering (interaction) energy of the atomic cloud.

Among the applications that take advantage of this lower-dimensional character, quantum interferometry protocols can be particularly sensitive to the true three-dimensional nature of the condensate. In recent work~\cite{tacla10}, simulations of a nonlinear BEC interferometer in highly elongated geometries showed significant deviations from the quasi-1D model with increasing strength of the nonlinear scattering interaction.  Similar effects have been shown to impact the propagation of solitons in quasi-1D attractive BECs~\cite{muryshev02,sinha06,khaykovich06}; this is a potential source of problems in implementations of matter-wave interferometers.

Here we study effects associated with the emergence of three-dimensional behavior in reduced-dimension BECs trapped by highly anisotropic potentials.  We develop a perturbative Schmidt decomposition of the condensate wave function between the tightly confined (transverse) direction(s) and the loosely confined (longitudinal) direction(s).  The perturbation theory is valid so long as the nonlinear scattering energy is small compared to the transverse energy scales.  In contrast to variational methods~\cite{das02a,das02b,gerbier04}, corrections to the reduced-dimension approximation are found without relying on any \textit{a priori\/} assumptions about the condensate wave function or the shape of the trapping potential.  Because the perturbation formalism is tied to the Schmidt decomposition, it automatically encodes information about the entanglement between the spatial and longitudinal directions.  The dominant Schmidt term corresponds to the optimal product-state approximation to the condensate wave function~\cite{lockhart02}; within this dominant term, the perturbation formalism provides corrections to the lowest-order transverse and longitudinal wave functions of the reduced-dimension approximation; the main effect is a reshaping of the BEC in the tightly confined direction as the strength of the nonlinear scattering interaction increases.  The next Schmidt term describes the lowest-order entanglement between the transverse and longitudinal directions; the perturbation formalism allows us to calculate the form and amount of this entanglement.

We compare the results of our perturbation theory with numerical integration of the three-dimensional Gross-Pitaevskii (GP) equation. We study the case of $^{87}$Rb condensates trapped by cigar-shaped potentials for various atom numbers and experimentally accessible parameters. By considering different trapping potentials, we also investigate the dependence of the results on the inhomogeneity of the longitudinal potential. We find surprisingly good agreement even for relatively high nonlinearities; in particular, even for such stronger nonlinearities, the entanglement between transverse and longitudinal directions remains remarkably small, which allows the condensate to be well described by a product wave function that corresponds to a single Schmidt~term.

We begin our discussion by briefly reviewing in Sec.~\ref{sec:lowDBECs} the mean-field description of quasi-reduced-dimension BECs, followed by the perturbative derivation of the Schmidt decomposition of the condensate wave function in \hbox{Sec.~\ref{sec:decomposition}}.  Comparison of our perturbative approximation with numerical solutions of the time-independent, three-dimensional GP equation is presented in Sec.~\ref{sec:simulations} for the parameters of $^{87}$Rb condensates trapped by different cigar-shaped potentials and various atom numbers.  The spatial entanglement between transverse and longitudinal directions is particularly analyzed in Sec.~\ref{sec:entanglement}.  Final remarks are given in \hbox{Sec.~\ref{sec:conclusion}}.


\section{Reduced-dimension approximation to a BEC in a highly anisotropic trap}
\label{sec:lowDBECs}
In the mean-field approximation, one describes a condensate of $N$ atoms at zero temperature by a wave function $\psi$ (normalized to unity) that is determined by the time-independent Gross-Pitaevskii equation
\begin{equation}
\label{timeindepGPE}
\mu\psi = \bigg(-\frac{\hbar^2}{2 M}\nabla^2 + V + g (N-1)|\psi|^2 \bigg)\psi\;,
\end{equation}
where $V$ is the external trapping potential, $\mu$ is the chemical potential, and $g = 4\pi\hbar^2a/M$ is the scattering strength determined by the $s$-wave scattering length $a$ and the atomic mass $M$.  For brevity, we generally use
\begin{equation}
\tilde g\equiv g(N-1)
\end{equation}
in the following.

In the case of highly anisotropic potentials, the atomic cloud is loosely trapped by a potential $V_L({\bm r})$ in $d$ dimensions, referred to as longitudinal ($L$) dimensions, as opposed to the remaining $D=3-d$ transverse degrees of freedom ($T$), which are tightly confined in a potential $V_T({\bm\rho})$. If the scattering interaction is sufficiently small compared to the transverse energy scale, one can neglect the effect of the nonlinear interaction on the atomic transverse degrees of freedom and hence approximate the condensate wave function by a product wave function,
\begin{equation}
\label{ansatz}
    \uppsi_0({\bm\rho},{\bm r}) = \xi_0({\bm \rho})\phi({\bm r})\;,
\end{equation}
where $\xi_0({\bm\rho})$ is the ground-state wave function of the bare transverse potential and $\phi({\bm r})$ is the solution of the $d$-dimensional, longitudinal GP equation
\begin{equation}
\label{dDGPE}
\left(-\frac{\hbar^{2}}{2 M}\nabla_L^2+
V_L({\bm r}) + \tilde g\eta_{T}|\phi({\bm r})|^{2}\right)\!\phi({\bm r}) = \mu_L \phi({\bm r})\;,
\end{equation}
which is found by plugging the product ansatz~(\ref{ansatz}) into the GP equation~(\ref{timeindepGPE}) and projecting the result onto the subspace spanned by $\xi_0$.
Here $\mu_L=\mu-E_0$ is the longitudinal part of the chemical potential, $E_0$ is the transverse ground-state energy, and
\begin{equation}
\eta_T=\int d^D\!\rho\,|\xi_0({\bm \rho})|^4\;.
\end{equation}

In this reduced-dimension approximation, the transverse and longitudinal degrees of freedom are decoupled, which should hold as long as the number of atoms in the condensate is small compared to an (upper) critical atom number $N_T$, defined as the number at which the scattering energy becomes comparable to the transverse kinetic energy, i.e.,
\begin{equation}
\label{NT}
     \frac{g}{2} (N_T-1) \eta =
     \frac{\hbar^2}{2M}\int d^D\!\rho\,|\nabla\xi_0|^2\;,
\end{equation}
where $\eta = \int d^3r\,|\psi|^4$ is a measure of the inverse volume occupied by the condensate wave function~$\psi$.

As $N$ approaches $N_T$, one can no longer neglect the effects of the scattering interaction on the condensate transverse degrees of freedom; as a result, the product ansatz~(\ref{ansatz}) is no longer a good approximation to the three-dimensional wave function.  Such effects are responsible not only for redefining the transverse and longitudinal wave functions, but also for entangling the spatial directions. We show below that these effects can be readily calculated in the perturbative regime where $N$ is small compared to $N_T$ by performing a perturbative Schmidt decomposition of the condensate wave function.


\section{Perturbative Schmidt decomposition of condensate wave function}
\label{sec:decomposition}

Our goal is to find an approximate solution to Eq.~(\ref{timeindepGPE}) that correctly accounts for the nonlinear effects on the tightly confined directions that are neglected by the reduced-dimension approximation.  Instead of proposing an alternative to the product ansatz~(\ref{ansatz}), we look for a solution to the GP equation in the form of the Schmidt decomposition,
\begin{eqnarray}
\label{psi}
  \psi({\bm \rho},{\bm r}) = \sum_{n=0}^\infty c_n \chi_n({\bm \rho}) \phi_n({\bm r})\;,
\end{eqnarray}
where $\{\chi_n\}$ and $\{\phi_n\}$ form orthonormal Schmidt bases in the transverse and longitudinal directions and the $c_n$'s are the (nonnegative) Schmidt coefficients (the squares $c_n^2$ are the eigenvalues of the marginal transverse and longitudinal density operators).  The decomposition~(\ref{psi}) is guaranteed to exist, but the Schmidt basis must be determined from the GP equation~(\ref{timeindepGPE}).  We can assume that the condensate wave function and the Schmidt basis functions are real.

As the deviations from the reduced-dimension approximation arise in a regime where the scattering interaction can be considered as a perturbation to the single-particle transverse Hamiltonian, we can solve for the Schmidt decomposition in successive orders of a perturbation theory.  We begin by writing the GP equation in the form
\begin{equation}
\label{3DGPE}
\mu\psi = \Big( H_{T} + \epsilon H_{L} + \epsilon \tilde g|\psi|^2 \Big)\psi\;,
\end{equation}
where $H_{T(L)}=-(\hbar^2/2M)\nabla_{T(L)}^2 + V_{T(L)}$ is the transverse (longitudinal) single-particle Hamiltonian and $\epsilon$ is a formal perturbation parameter.  Notice that due to the asymmetry of the trapping potential, the longitudinal Hamiltonian $H_L$ and the nonlinear scattering interaction are treated as of the same size, both being order $\epsilon$ smaller than the transverse Hamiltonian.

As $\epsilon$ goes to zero, we expect the solution of the GP equation~(\ref{timeindepGPE}) to reduce to the product wave function~(\ref{ansatz}), in which the Schmidt decomposition has only one term.  As $\epsilon$ increases, the Schmidt decomposition acquires additional terms.  We are thus motivated to treat the Schmidt decomposition formally as a power-series expansion in $\epsilon$:
\begin{eqnarray}
\label{Schmidt}
  \psi=\sum_{n=0}^\infty \epsilon^n\chi_n\phi_n\;.
\end{eqnarray}
In developing the perturbation theory, we find it convenient to absorb the Schmidt coefficients into the transverse Schmidt basis functions, which thus satisfy the orthogonality relation
\begin{equation}\label{chinchim}
\langle\chi_n|\chi_m\rangle=c_n^2\delta_{nm}\;.
\end{equation}
The longitudinal Schmidt basis functions are orthonormal,
\begin{equation}\label{phinphim}
\langle\phi_n|\phi_m\rangle=\delta_{nm}\;.
\end{equation}
The consequences of the normalization of the overall wave function and of the orthogonality relations~(\ref{chinchim}) and~(\ref{phinphim}) are spelled out in Appendix~\ref{ap:expansion}.

We also look for the chemical potential as an expansion in powers of $\epsilon$,
\begin{equation}
\mu = \sum_{m=0}^\infty \epsilon^m\mu_m\;,
\end{equation}
and similarly for the Schmidt basis functions as
\begin{align}\label{expandchin}
\chi_n &= \sum_{m=0}^\infty \epsilon^m \chi_{nm}\;,\\
\phi_n &= \sum_{m=0}^\infty \epsilon^m \phi_{nm}\;.
\label{expandphin}
\end{align}

We seek a solution to the 3D~GP equation~(\ref{3DGPE}) to first order in $\epsilon$.  Thus we are looking for a solution of the form
\begin{align}
\label{psiSch}
\uppsi_1({\bm \rho},{\bm r})
= &[\chi_{00}({\bm \rho})+\epsilon\chi_{01}(\bm\rho)][\phi_{00}({\bm r})+\epsilon \phi_{01}(\bm r)]\nonumber\\
&+ \epsilon\chi_{10}({\bm \rho})\phi_{10}({\bm r})
\;,
\end{align}
which is spatially entangled (or nonseparable).  We relegate the details of the straightforward, but tedious derivation of the perturbative equations to Appendix~\ref{ap:expansion} and only present the results here.

The $m=0$ terms in the $n=0$ Schmidt term correspond, as expected, to the idealized description of a quasi-$d$-dimensional BEC summarized in Sec.~\ref{sec:lowDBECs}: $\chi_{00}=\xi_0$ is the ground-state wave function of the bare transverse potential, and this means that $\mu_0=E_0$ is the ground-state energy of the transverse trap; $\phi_{00}$ is determined by the reduced-dimension GP equation~(\ref{dDGPE}), here written as
\begin{equation}
	\mu_1 \phi_{00} = (H_L + \tilde g\eta_{T} \phi_{00}^2) \phi_{00}\;,\label{phi00}
\end{equation}
with the nonlinear interaction renormalized by the average of $\chi_{00}^2$ over itself,
\begin{equation}\label{etaT}
\eta_{T}\equiv\langle \chi_{00} | \chi_{00}^3 \rangle=\int d^D\!\rho\,\chi^4_{00}({\bm \rho})\;.
\end{equation}
Hereafter, for brevity, we usually represent spatial integrals in terms of bra-ket inner products.
The longitudinal GP equation~(\ref{phi00}) also determines the first correction, $\mu_1$, to the chemical potential.

There are four first-order corrections to be calculated.  The functions $\chi_{01}$ and $\phi_{01}$ are the first corrections within the $n=0$ Schmidt term, i.e., to the transverse and longitudinal wave functions $\chi_{00}$ and $\phi_{00}$, whereas $\chi_{10}$ and $\phi_{10}$ describe the lowest-order spatial entanglement.  The transverse functions $\chi_{01}$ and $\chi_{10}$ are determined by the linear differential equations
\begin{equation}\label{DEchi01chi10}
(\mu_0 - H_T)\frac{\chi_{01}}{\eta_L}
=\tilde g(\chi_{00}^2-\eta_T)\chi_{00}=
(\mu_0 - H_T)\frac{\chi_{10}}{\Delta\eta_L}\;,
\end{equation}
where
\begin{equation}\label{etaL}
\eta_{L}\equiv \langle \phi_{00} | \phi_{00}^3 \rangle=\int d^d r\,\phi^4_{00}({\bm r})
\end{equation}
is the average of the probability distribution $\phi_{00}^2$ over itself, and
\begin{equation}\label{Deltaeta}
\Delta\eta_L^{2}\equiv\langle \phi_{00}^3 |\phi_{00}^3 \rangle - \eta_{L}^2\ge0
\end{equation}
is the variance of $\phi_{00}^2$.  Notice that the first-order transverse corrections, $\chi_{01}$ and $\chi_{10}$, are driven by inhomogeneities in the zero-order transverse profile $\chi_{00}^2$.  We can write a solution of Eq.~(\ref{DEchi01chi10}) in terms of the eigenfunctions and eigenenergies of the transverse Hamiltonian $H_T$, i.e., $H_T\xi_n=E_n\xi_n$,
\begin{equation}\label{chi01chi10}
\frac{\chi_{01}}{\eta_L}=\frac{\chi_{10}}{\Delta\eta_L}
=-\tilde g\sum_{n=1}^\infty \xi_n\frac{\langle\xi_n|\xi_0^3\rangle}{E_n-\mu_0}\;.
\end{equation}

The longitudinal function $\phi_{01}$ is determined by the equation
\begin{equation}\label{phi01}
(\mu_1-H_L-3\tilde g\eta_T\phi_{00}^2)\phi_{01}
=-3\tilde g^2\UpsilonT\phi_{00}^5-\mu_2\phi_{00}\;.
\end{equation}
Here
\begin{equation}\label{UpsilonT}
\UpsilonT\equiv
\sum_{n=1}^\infty\frac{\langle\xi_n|\xi_0^3\rangle^2}{E_n-\mu_0}\ge0
\end{equation}
is a coupling parameter, which is determined solely by the properties of the transverse trap
and which characterizes the strength of the coupling of transverse and longitudinal directions; the quantity $\eta_T^2/\UpsilonT$ can be thought of as the relevant quantification of the transverse energy scale as far as the perturbation theory is concerned.  Projecting Eq.~(\ref{phi01}) onto $\phi_{00}$ gives an expression for the correction to the chemical potential,
\begin{equation}\label{mu2}
\mu_2
=2\tilde g\eta_T\langle\phi_{01}|\phi_{00}^3\rangle
-3\tilde g^2(\eta_L^2+\Delta\eta_L^2)\UpsilonT\;,
\end{equation}
which shows that Eq.~(\ref{phi01}) is a linear integro-differential equation for $\phi_{01}$.

The remaining longitudinal function, $\phi_{10}$, is given by a trivial algebraic equation,
\begin{equation}	
	\phi_{10} = \frac{\phi_{00}^2 - \eta_{L}}{\Delta\eta_L}\phi_{00}\;.
\label{phi10}
\end{equation}
The first-order longitudinal corrections are driven by inhomogeneities in the zero-order longitudinal profile $\phi_{00}^2$.

It is an easy matter to derive from Eqs.~(\ref{chi01chi10}) and (\ref{phi10}) that
\begin{align}\label{chi01chi003}
\frac{\langle\chi_{01}|\chi_{00}^3\rangle}{\eta_L}
=\frac{\langle\chi_{10}|\chi_{00}^3\rangle}{\Delta\eta_L}
&=-\tilde g\sum_{n=1}^\infty\frac{\langle\xi_n|\xi_0^3\rangle^2}{E_n-\mu_0}
=-\tilde g\UpsilonT\;,\\
\langle\phi_{10}|\phi_{00}^3\rangle&=\Delta\eta_L\;.
\label{phi01phi003}
\end{align}
At the order we are working, the first two Schmidt coefficients are given by $c_0^2=\langle\chi_{00}|\chi_{00}\rangle=1$ and
\begin{equation}
\label{c1}
	c_1^2 = \langle\chi_{10}|\chi_{10}\rangle=
    \tilde g^2 \Delta\eta_L^2
    \sum_{n=1}^\infty\frac{\langle \xi_{n} | \chi_{00}^3\rangle^2}{(E_n - \mu_0)^2}.
\end{equation}

The only nonlinear equation we have to solve is the (differential) longitudinal GP equation~(\ref{phi00}) for $\phi_{00}$, but we are faced with solving the linear integro-differential equation~(\ref{phi01}) for $\phi_{01}$.  It is easier and more instructive to combine these two equations into a single nonlinear differential equation for the longitudinal contribution to the $n=0$ Schmidt term,
\begin{equation}
\phi_0=\phi_{00}+\epsilon\phi_{01}+O(\epsilon^2)\;.
\end{equation}
To do this, we write Eqs.~(\ref{phi00}) and~(\ref{phi01}) in the forms
\begin{align}
0&=(\mu_1-H_L)\phi_{00}
-\tilde g\eta_T\phi_0^3+3\epsilon\tilde g\eta_T\phi_0^2\phi_{01}+O(\epsilon^2)\;,\\
0&=(\mu_1-H_L)\epsilon\phi_{01}-3\epsilon\tilde g\eta_T\phi_0^2\phi_{01}\nonumber\\
&\phantom{=(\mu_1}+3\epsilon\tilde g^2\UpsilonT\phi_{0}^5+\epsilon\mu_2\phi_{0}+O(\epsilon^2)\;.
\end{align}
Identifying $\tilde{\mu}_L=\mu_1+\epsilon\mu_2$ as the longitudinal part of the chemical potential, we can add these two equations (and then set $\epsilon=1$) to obtain a GP-like equation for $\phi_0$, accurate to first order in $\epsilon$:
\begin{align}
\label{effGPE}
	\tilde{\mu}_L \phi_0
    &= (H_L + \tilde g \eta_{T} \phi_0^2 - 3\tilde g^2\UpsilonT\phi_{0}^4)\phi_0\nonumber\\
    &= \bigl[ H_L + g \eta_{T} (N-1) \phi_0^2 - 3g^2\UpsilonT(N-1)^2\phi_{0}^4\bigr]\phi_0\;.
\end{align}
In the second form of the right-hand side, we restore the $N$ dependence to reveal the intrinsic coupling strengths.  Relative to a GP equation, this longitudinal equation has an additional quintic term, which acts as an effective three-body, attractive interaction among the atoms.  This attractive interaction is mediated by the changes in the transverse wave function, as evidenced by the appearance of the (nonnegative) coupling parameter $\UpsilonT$ in the coupling strength $3g^2\UpsilonT$.  Such a self-focusing interaction has also been used to study the propagation of solitons in attractive quasi-1D condensates~\cite{muryshev02,sinha06,khaykovich06}.

The coupling constants in Eq.~(\ref{effGPE}) can be calculated explicitly for a transverse harmonic potential, which we use henceforth.  In this case, the transverse ground-state wave function is the Gaussian
\begin{equation}
\label{gaussian}
	\chi_{00}({\bm \rho}) = \frac{e^{-\rho^2/2\rho_0^2}}{(\pi\rho_0^2)^{D/4}}\;,
\end{equation}
where $\rho_0 = \sqrt{\hbar/M\omega_T}$.  It is easy to see that
\begin{equation}
\label{etaTD}
	\eta_{T} = \left(\frac{1}{\sqrt{2\pi}\rho_0}\right)^{D}\;.
\end{equation}
Moreover, for a pancake ($D=1$), we have
\begin{equation}\label{overlap1D}
\langle \xi_{n} | \xi_0^3 \rangle=
\begin{cases}
\displaystyle{\frac{(-1)^{n/2}}{\sqrt{\pi\, n!}}\eta_T \Gamma\left(\frac{n + 1}{2}\right)}\;,
&\mbox{$n$ even,}\\
0\;,&\mbox{$n$ odd.}
\end{cases}
\end{equation}
For a cigar ($D=2$), if we use polar co\"ordinates for the transverse eigenfunctions, they take the form $\xi_{n_r m}(\rho,\varphi)$, with $n_{r}$ and $m$ being radial and azimuthal quantum numbers and with the eigenenergies given by $E_{n_r m}=\hbar\omega_T(2n_r+|m|+1)$.  Then we find that
\begin{equation}
\langle \xi_{n_{r}m} | \xi_{00}^3 \rangle=2^{-n_{r}}\eta_T\delta_{m0}\;.\label{overlap2D}
\end{equation}
It follows from Eq.~(\ref{UpsilonT}) that $\UpsilonT$ is given by
\begin{equation}
\UpsilonT=
\begin{cases}
\displaystyle{\vphantom{\Bigg(\biggr)}\frac{\eta_T^2}{\hbar\omega_T}\ln(8-4\sqrt{3})\;,}&\mbox{$D=1$ (pancake),}\\
\displaystyle{\vphantom{\Bigg(\biggr)}\frac{\eta_T^2}{2\hbar\omega_T}\ln\frac{4}{3}\;,}&\mbox{$D = 2$ (cigar).}
\end{cases}
\end{equation}
As a result, the coupling constants for a cigar-shaped trap ($d=1$) are $g \eta_T = 2 \hbar\omega_T a$ and $3g^2\UpsilonT = 6\hbar\omega_T a^2\ln(4/3)$, whereas for a quasi-2D pancake ($d=2$), we obtain $g \eta_T = 2\sqrt{2\pi}\hbar\omega_T \rho_0 a$ and $3g^2\UpsilonT = 24\pi\hbar\omega_T \rho_0^2a^2\ln(8-4\sqrt{3})$.


\section{Numerical Simulations}
\label{sec:simulations}

The accuracy of the perturbative Schmidt decomposition of the condensate wave function can be checked by direct comparison to the numerical solution of the three-dimensional GP equation~(\ref{timeindepGPE}) for various atom numbers.

\subsection{Trap geometry and numerical integrations}
We restrict our comparisons to the case of highly elongated (cigar-shaped) condensates of $^{87}$Rb atoms in the $|F = 1, m_F = -1\rangle$ hyperfine state, for which $a=100.4\,a_0$, with $a_0$ being the Bohr radius, trapped by potentials of the form
\begin{equation}
\label{potential}
    V(\rho,z) = \frac{1}{2}(M\omega_T^2 \rho^2 + k z^q)\;,
\end{equation}
with $q = 2, 4,$ and $10$.  Such choice of potentials allows us to explore how the hardness of the potentials affects the results.  A hard-walled longitudinal trap corresponds to the limit $q\rightarrow \infty$.  We set the transverse frequency to $350\,$Hz, which gives $\rho_0\simeq 0.6\,\mu$m.

For the case of a harmonic longitudinal trap ($q=2$), we set the longitudinal frequency to $3.5\,$Hz and find that $N_T \simeq 14\,000$ atoms~\cite{NT}.  To compare the simulations for the different longitudinal power-law potentials, we choose the stiffness parameter $k$ so that $N_T$ has the same value for the two other values of $q$; thus all the traps have the same one-dimensional regime of atom numbers.  We define $z_0\equiv(\hbar^2/Mk)^{1/(q+2)}$ as a measure of the bare ground-state width in the longitudinal direction ($z_0$ simplifies to the analog of $\rho_0$ for a harmonic longitudinal trap).  With these choices, the aspect ratio of the bare traps, $\rho_0\!:\!z_0$, is approximately equal to $1\!:\!10$, $1\!:\!24$, and $1\!:\!57$ for $q=2$, 4, and 10.

In addition to integrating the 3D~GP equation~(\ref{timeindepGPE}), we also numerically integrate the quasi-1D GP equation~(\ref{phi00}) and the perturbative quintic equation~(\ref{effGPE}) for the trapping potentials~(\ref{potential}) and different atom numbers~\cite{numerics}.  The latter two integrations yield the longitudinal wave functions $\phi_{00}(z)$ and $\phi_0(z)$ and also determine $\mu_1$ and $\tilde{\mu}_L$.

Given $\phi_{00}(z)$, we can determine the remaining Schmidt functions: $\phi_{10}$ follows trivially from Eq.~(\ref{phi10}), whereas the transverse terms can be found from Eqs.~(\ref{chi01chi10}) and~(\ref{overlap2D}), together with $g\eta_T/\hbar\omega_T=2a$, all of which yields
\begin{align}
\chi_{0}(\rho)
&= \xi_{00}(\rho)- a \eta_{L} (N-1) \sum_{n_{r}=1}^\infty \frac{\xi_{n_{r}0}(\rho)}{2^{n_{r}} n_{r}}\;,\label{chi0} \\
\chi_{10}(\rho)
&= - a \Delta\eta_L (N-1) \sum_{n_{r}=1}^\infty \frac{\xi_{n_{r}0}(\rho)}{2^{n_{r}} n_{r}}\;,
\label{chi1}
\end{align}
where the functions $\xi_{n_{r}0}(\rho)=e^{-\rho^2/2 \rho_0^2} L_{n_{r}}(\rho^2/\rho_0^2)/\sqrt{\pi}\rho_0$ are the $m=0$ (azimuthally symmetric) Laguerre-Gaussian eigenfunctions for the two-dimensional harmonic potential, with energies $E_{n_r0}=2\hbar\omega_T n_r$.

From the normalization of Eq.~(\ref{chi1}), we get
\begin{equation}
\label{c1harmonic}
	c_1 = \sqrt{\langle\chi_{10}|\chi_{10}\rangle}=
    \sqrt{{\rm Li}_{2}(1/4)} (N-1) a\Delta\eta_L\;,
\end{equation}
where we use the polylogarithm function ${\rm Li}_{s}(z)\equiv\sum_{n=1}^\infty z^n/n^s$.

\begin{figure}
\includegraphics[scale=.4]{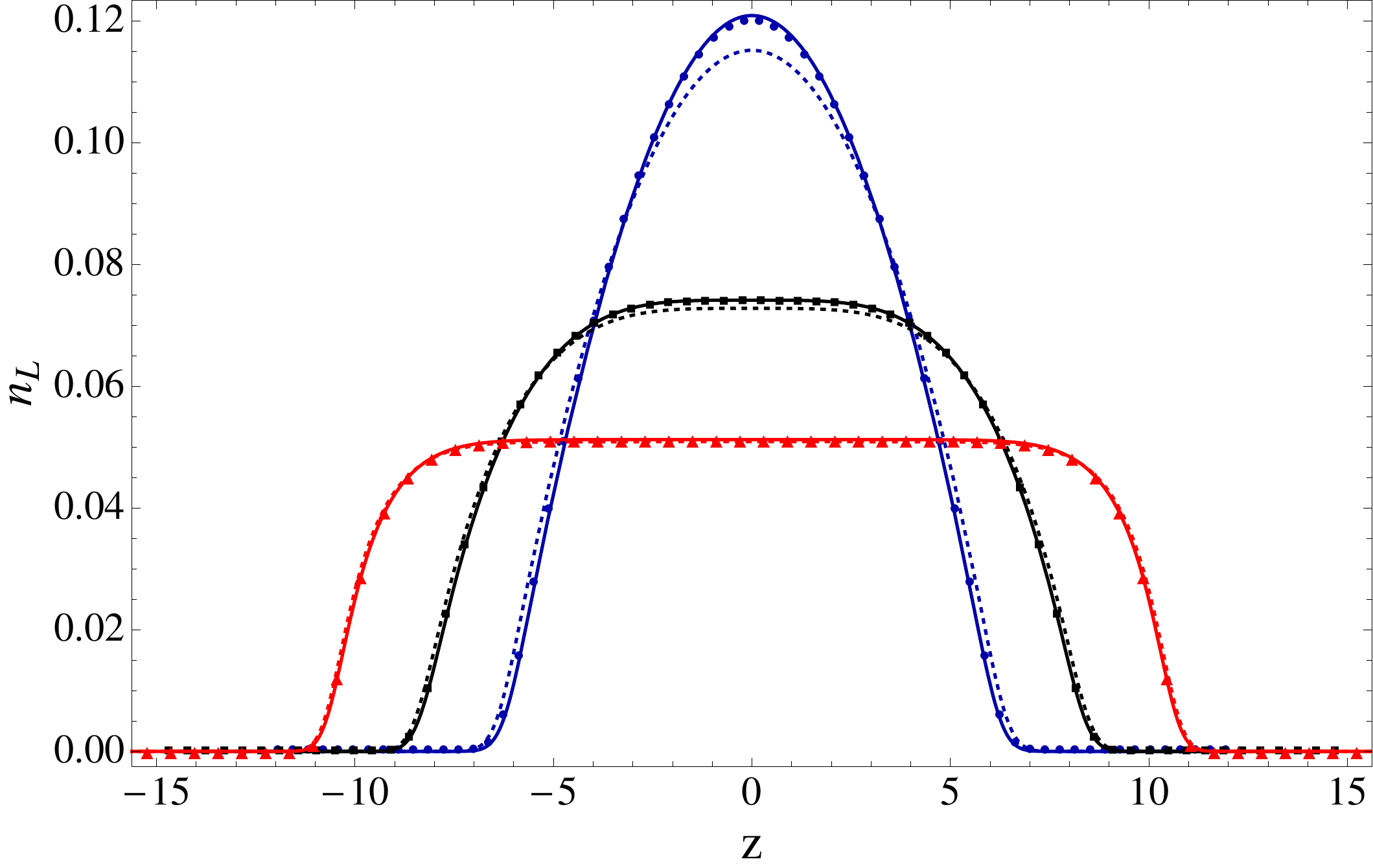}
\caption{(Color online) Longitudinal marginal distribution (in harmonic trap units) for a condensate of 1\,000 atoms. The discrete points are the results of the integration over the transverse plane of the 3D~GP ground-state solution for different trap geometries: circles (blue) signify  $q=2$, squares (black) $q=4$, and triangles (red) $q=10$. The corresponding solid lines represent the distribution given by the $n=0$ longitudinal Schmidt function, $|\phi_0(z)|^2$, whereas the dotted lines show the unperturbed distribution $|\phi_{00}(z)|^2$.  The marginal distribution is well described by the Schmidt function for all values of $q$.  In the case of a harmonic trap ($q=2$), the effect of the quintic coupling on $|\phi_0(z)|^2$ is evident, but becomes less pronounced for higher $q$, as the longitudinal distribution becomes more homogeneous.}
\label{fig:nL}
\end{figure}

\subsection{Dominant Schmidt term}

In this subsection, we study the dominant ($n=0$) Schmidt term, using various quantities to compare the predictions of the Schmidt perturbation theory with the numerical predictions of the 3D~GP equation.  We also include the predictions of the quasi-1D, reduced-dimension approximation to determine how significant the perturbative Schmidt terms are.

By integrating out the transverse dimensions from the numerical solution of the 3D~GP equation, we calculate the axial (longitudinal) marginal distribution
\begin{equation}
n_L(z) = \int d^2\!\rho\,|\psi(\rho,z)|^2\;.
\end{equation}
At the order we are working in perturbation theory, this marginal distribution is given, according to Eq.~(\ref{psiSch}), by $|\phi_0(z)|^2$, as the contribution from the $n=1$ Schmidt term is of higher order.  In the quasi-1D approximation, this marginal distribution is given by $|\phi_{00}(z)|^2$.

\begin{figure}
\includegraphics[scale=.4]{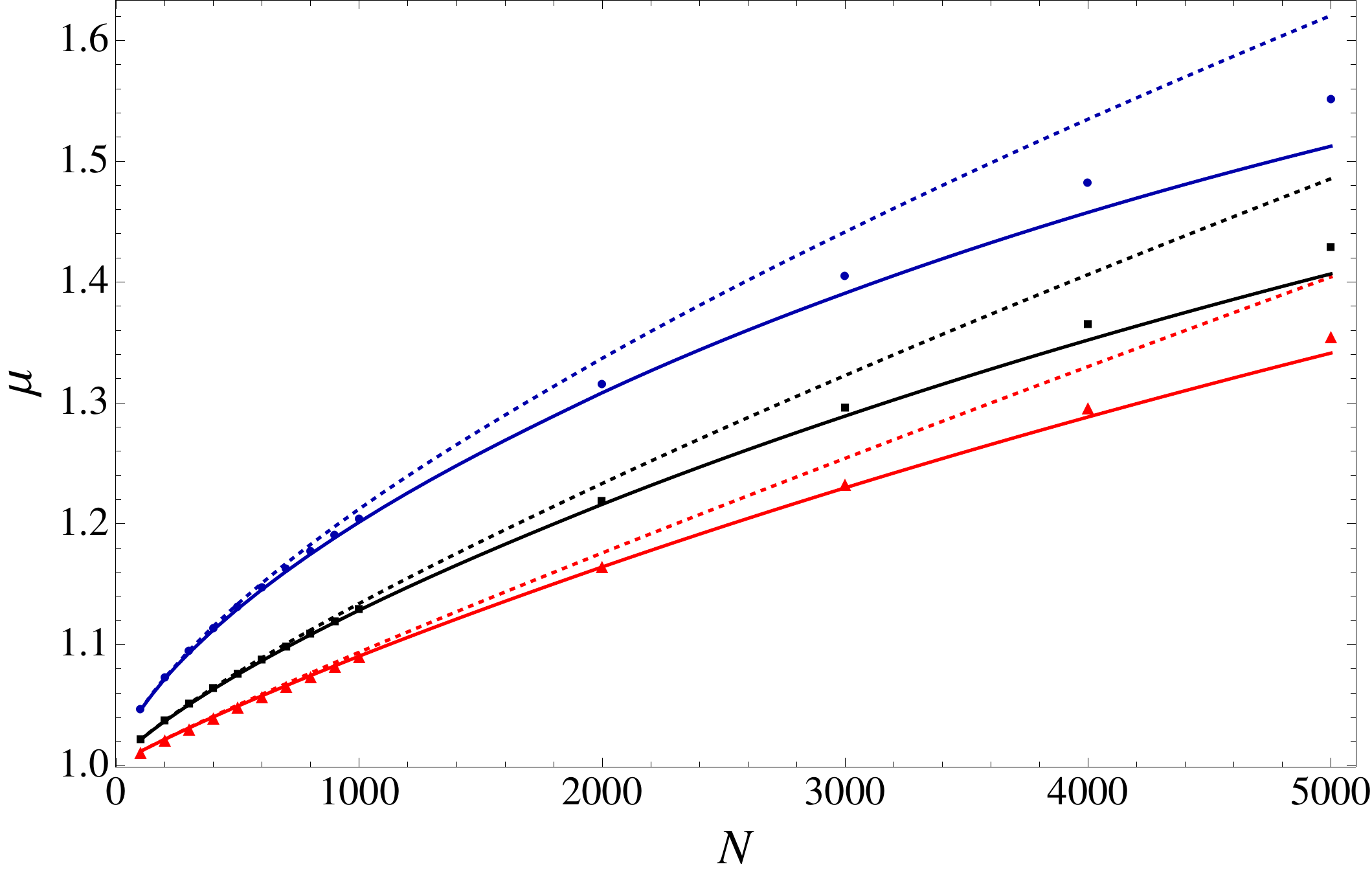}
\caption{(Color online) Chemical potential in units of $\hbar\omega_T$ as a function of the number of atoms in the condensate for longitudinal potentials~(\ref{potential}).  Discrete points represent the result obtained from the numerical integration of the 3D~GP equation, with circles (blue) for $q = 2$, squares (black) for $q = 4$, and triangles (red) for $q = 10$.  The solid lines give the corresponding approximation coming from the perturbative quintic equation~(\ref{effGPE}), whereas the dotted lines are the estimates from the quasi-1D GP equation~(\ref{phi00}).  The correction introduced by the perturbation theory becomes quite significant as $N$ increases above 1\,000 and does indeed lead to a better approximation of the 3D~numerical results in comparison with the quasi-1D model, although the perturbation theory is noticeably failing, even for $q=10$, as $N$ approaches~5\,000.}
\label{fig:mu}
\end{figure}

In Fig.~\ref{fig:nL}, we plot $n_L(z)$ against the distributions $|\phi_0(z)|^2$ and $|\phi_{00}(z)|^2$ for a condensate of 1\,000 atoms and the three different values of $q$.  The marginal distribution is very well described by the $n=0$ Schmidt function $\phi_0(z)$ for all the potentials~(\ref{potential}). The $q=2$ case is particularly interesting, for the effect of the correction provided by the quintic coupling in Eq.~(\ref{effGPE}) proves to be quite noticeable due to the inhomogeneity of the harmonic potential.  As $q$ increases, however, this correction becomes less important, because the axial distribution becomes more homogenous as a result of the more hard-walled and flat-bottomed potentials.

\begin{figure}
\includegraphics[scale=.4]{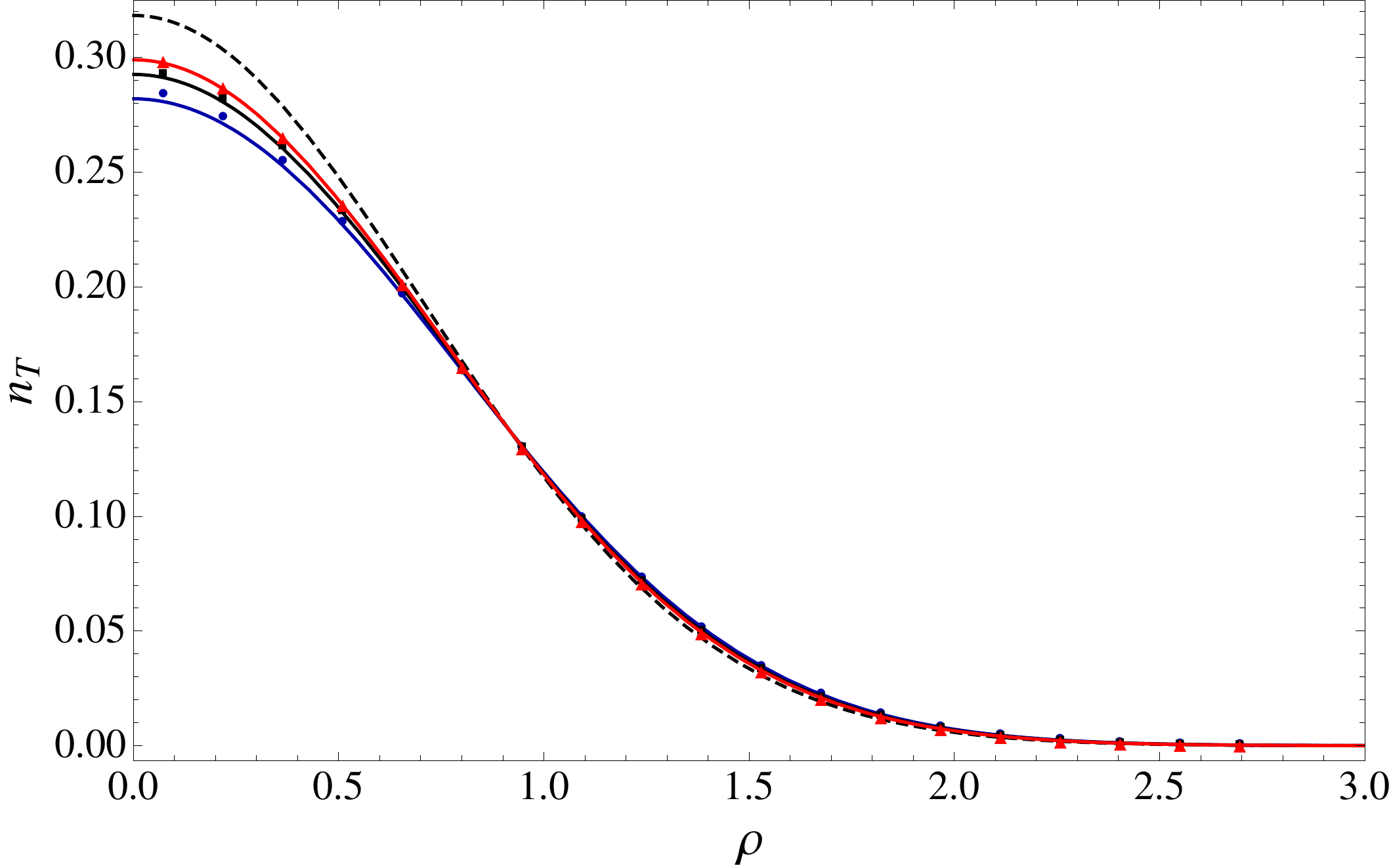}
\caption{(Color online) Transverse marginal distribution for a condensate of 1\,000 atoms in harmonic trap units.  The discrete points are the results of the integration over the axial axis of the 3D~GP ground-state solution for different trap geometries: circles (blue) denote $q=2$, squares (black) $q=4$, and triangles (red) $q=10$.  The corresponding solid lines represent the distribution $|\chi_0(\rho)|^2$, whereas the black dashed line represents the bare (Gaussian) distribution $|\xi_{00}(\rho)|^2$.  For all values of $q$, the marginal distribution is well described by the Schmidt function $|\chi_0(\rho)|^2$, which represents a major improvement over the bare distribution $|\xi_{00}(\rho)|^2$.}
\label{fig:nT}
\end{figure}

The better performance of the Schmidt function $\phi_0(z)$ over $\phi_{00}(z)$ is also evident in  predictions for the chemical potential $\mu$ as a function of the number of atoms in the condensate.  According to our perturbative expansion, the chemical potential is estimated to be $\tilde{\mu}=\hbar\omega_T + \tilde{\mu}_L=\hbar\omega_T+\mu_1+\mu_2$, as opposed to the estimate of the quasi-1D approximation, $\mu_{1{\rm D}}=\hbar\omega_T + \mu_1$; these only differ by the longitudinal contribution $\mu_2$.  The difference comes directly from the difference between the quasi-1D GP equation~(\ref{phi00}) and the perturbative quintic equation~(\ref{effGPE}), and it is from integrating these two equations that we get $\mu_1$ and $\tilde{\mu}_L$.  We compare, in Fig.~\ref{fig:mu}, the two approximations against the chemical potential given by the numerical integration of the 3D~GP equation.  Deviations from the quasi-1D model are again well captured by the perturbation theory, especially for potentials with higher values of~$q$.

We can also integrate out the longitudinal dimension to calculate the transverse (radial) marginal distribution $n_T(\rho) = \int dz\,|\psi(\rho,z)|^2$ and compare it with the approximate transverse distributions $|\chi_0(\rho)|^2$ and $|\xi_{00}(\rho)|^2$.  As shown in Fig.~\ref{fig:nT}, the transverse distribution for a condensate of 1\,000 atoms is well described by the Schmidt perturbation theory for all the potentials~(\ref{potential}); the perturbation theory correctly accounts for the spreading of the condensate in the radial direction.  In contrast to the axial profile in Fig.~\ref{fig:nL}, the correction to the bare radial distribution is not affected by the inhomogeneity of the longitudinal potentials.  Instead, as predicted by Eq.~(\ref{chi0}), it is set by the length ratio $a\eta_L$; the radial distributions are nearly the same because $\eta_L$ varies only slightly among the three values of~$q$.

\begin{figure}
\includegraphics[scale=.4]{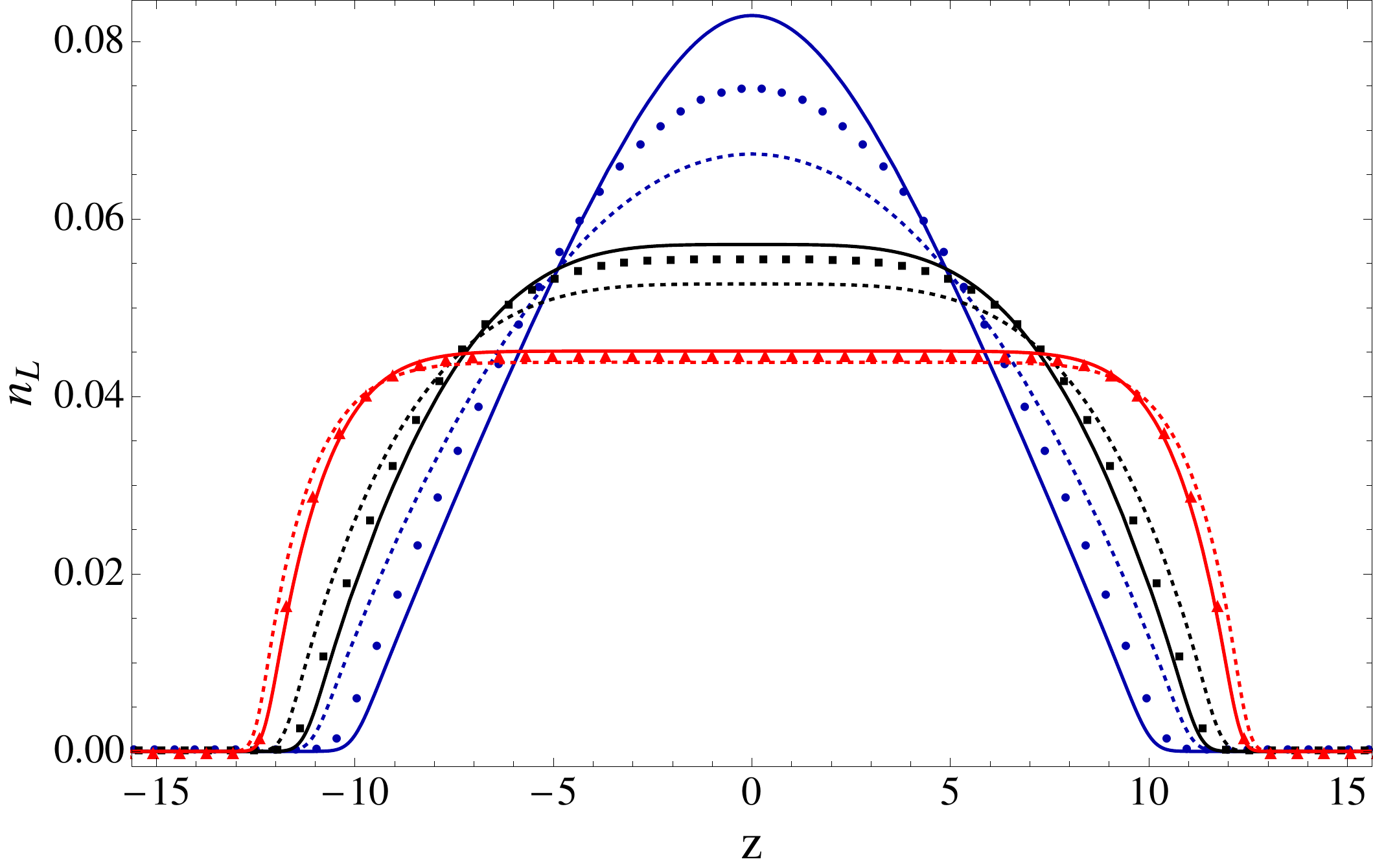}
\caption{(Color online) Longitudinal marginal distribution (in harmonic trap units) for a condensate of 5\,000 atoms.  The discrete points are the results of the integration over the transverse plane of the 3D~GP ground-state solution for different trap geometries: circles (blue) designate $q=2$, squares (black) $q=4$, and triangles (red) $q=10$.  The corresponding solid lines represent the distribution given by $|\phi_0(z)|^2$, whereas the dotted lines show the unperturbed distribution $|\phi_{00}(z)|^2$.  For this many atoms, the self-focusing quintic term clearly over-corrects the unperturbed distribution, signaling the breakdown of the perturbation theory.  This over-correction is especially evident for $q=2$.  For $q=4$ and $q=10$, the over-correction is not as bad, and the perturbation theory does a reasonably good job, even for this quite large number of atoms.}
\label{fig:nL5000}
\end{figure}

\begin{figure}
\includegraphics[scale=.4]{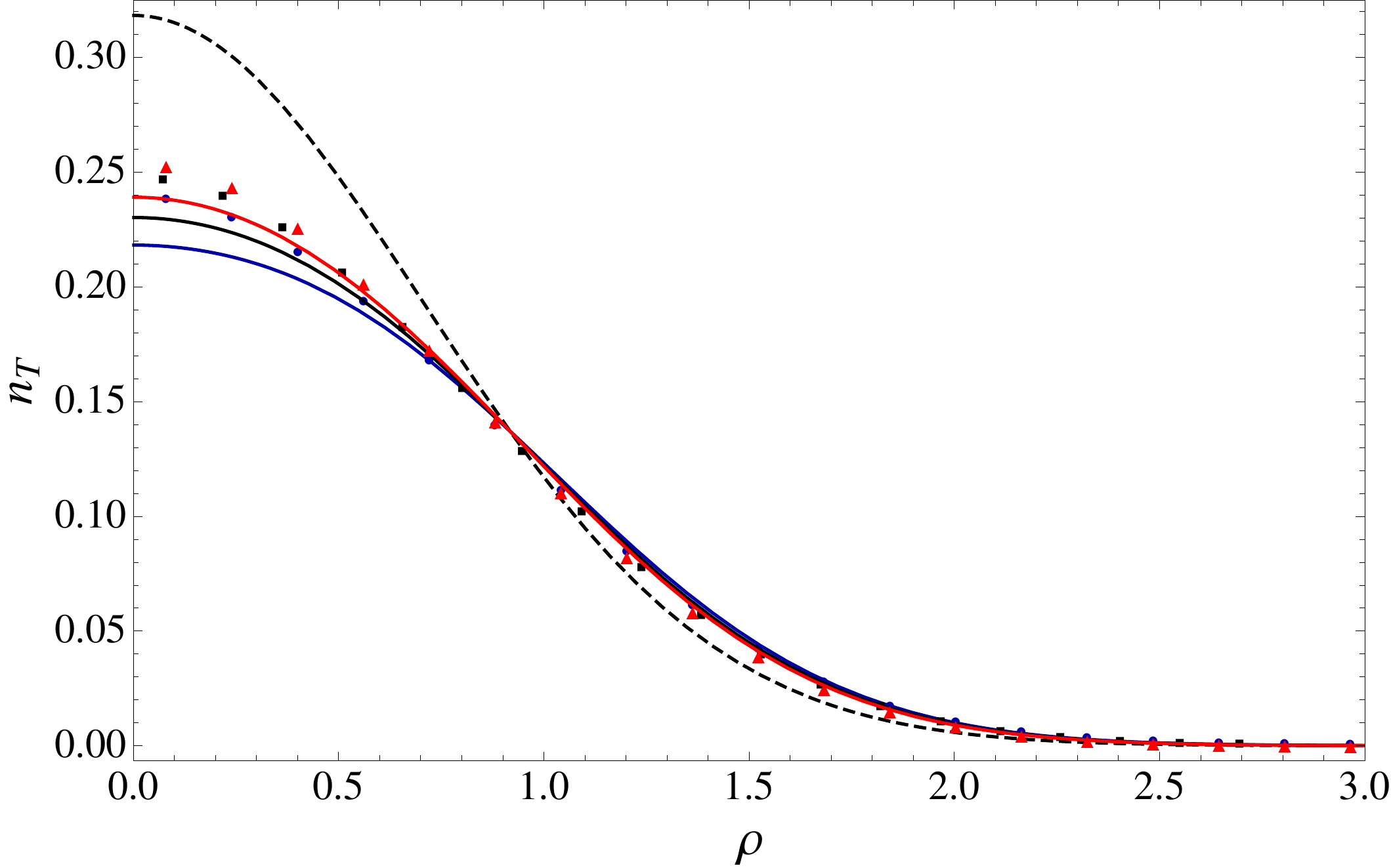}
\caption{(Color online) Transverse marginal distribution for a condensate of 5\,000 atoms in harmonic trap units.  The discrete points are the results of the integration over the axial axis of the 3D~GP ground-state solution for different trap geometries: circles (blue) denote $q=2$, squares (black) $q=4$, and triangles (red) $q=10$.  The corresponding solid lines represent the distribution $|\chi_0(\rho)|^2$, whereas the black dashed line represents the bare (Gaussian) distribution $|\xi_{00}(\rho)|^2$.  For this many atoms the perturbation theory starts to break down, as the Schmidt function $|\chi_0(\rho)|^2$ over-corrects the unperturbed distribution $|\xi_{00}(\rho)|^2$.  For all values of $q$, the marginal distribution is still reasonably well described by the perturbation theory.}
\label{fig:nT5000}
\end{figure}

The success of our perturbation theory emboldens us to push it a bit beyond where it really should work.  Figure~\ref{fig:nL5000} plots the longitudinal marginal distribution for 5\,000 atoms, and Fig.~\ref{fig:nT5000} plots the transverse marginal distribution for 5\,000 atoms.  In both figures, we can see the breakdown of the perturbation theory, although it performs surprisingly well, especially for higher values of $q$, given that this atom number is more than a third of the way to $N_T$.

\subsection{Spatial entanglement}
\label{sec:entanglement}

The last two Schmidt functions, $\chi_{10}(\rho)$ and $\phi_{10}(z)$, introduce the final correction to the quasi-1D, reduced-dimension approximation, namely, to the assumption of a spatially separable three-dimensional wave function.  The validity and importance of these corrections can be assessed in terms of two quantities.  The first of these is the probability $P_D$ of finding the numerically determined solution of the 3D~GP equation~(\ref{3DGPE}) outside the subspace spanned by the Schmidt wave functions $\chi_0\phi_0$ and $\chi_1\phi_1$.  This probability deficit tells us the extent to which the exact solution is confined to the two-dimensional subspace of the perturbative Schmidt wave function~(\ref{psiSch}) and is given by
\begin{equation}
\label{P}
	P_D= 1 - \tilde{c}_0^2 - \tilde{c}_1^2\;,
\end{equation}
where $\tilde{c}_0$ and $\tilde{c}_1$ are Schmidt-like coefficients obtained by projecting the exact 3D solution onto the (normalized) Schmidt basis functions, $\chi_0\phi_0/c_0$ and $\chi_1\phi_1/c_1$.  Computed values of the deficit $P_D$ for the different longitudinal potentials are displayed in Table~\ref{tab:probability} for various atom numbers.  The very small values of the deficit indicate the success of our perturbation theory.  For larger atom numbers, however, the population outside the two-dimensional space increases as the perturbation theory begins to break down.

\begin{table}
	\caption{\label{tab:probability}Probability deficit $P_D \times 10^{4}$.  Small values are indicative of the overall success of the two-Schmidt-term perturbation theory.}
	\begin{ruledtabular}
		\begin{tabular}{l|ccccc}
			$q$ & $N=1\,000$ & $2\,000$ & $3\,000$ & $4\,000$ & $5\,000$ \\
			 \hline
			2 & 0.19 & 1.76 & 6.88 & 18.05 & 39.23 \\
			4 & 0.03 & 0.44 & 1.81 & 4.91 & 10.41 \\
			10 & 0.05 & 0.21 & 0.88 & 1.44 & 3.06 \\
		\end{tabular}
	\end{ruledtabular}
\end{table}

Within this two-dimensional subspace, we can assess the validity of the perturbative wave function in terms of an entanglement measure.  Since the perturbative wave function has the form of a two-qubit entangled state, we can use Wootters's concurrence~\cite{wootters98} for a pair of qubits as the entanglement measure.

The concurrence of a bipartite pure state $|\Psi_{AB}\rangle$ varies smoothly from 0 for product states to 1 for maximally entangled states.  From its definition, $C = |\langle \Psi_{AB}^* | \sigma_y\otimes\sigma_y | \Psi_{AB}^* \rangle|$, in terms of the Pauli matrix $\sigma_y$ and the complex conjugate of $|\Psi_{AB}\rangle$, it is easy to show that the concurrence for the perturbative condensate wave function~(\ref{psiSch}) is given by
\begin{equation}
\label{concurrence}
	C = 2 c_0 c_1 = 2 \sqrt{{\rm Li}_{2}(1/4)}(N-1)a\Delta\eta_L\;,
\end{equation}
where we use the perturbative coefficients $c_0=1$ and $c_1$ as given by Eq.~(\ref{c1harmonic}).  The concurrence tells us about the amount of entanglement generated by the nonlinear interaction between the radial and axial directions.  This information, which quantifies the importance of the nonseparable corrections, is essentially contained in the Schmidt coefficient $c_1$.

\begin{figure}
\includegraphics[scale=.4]{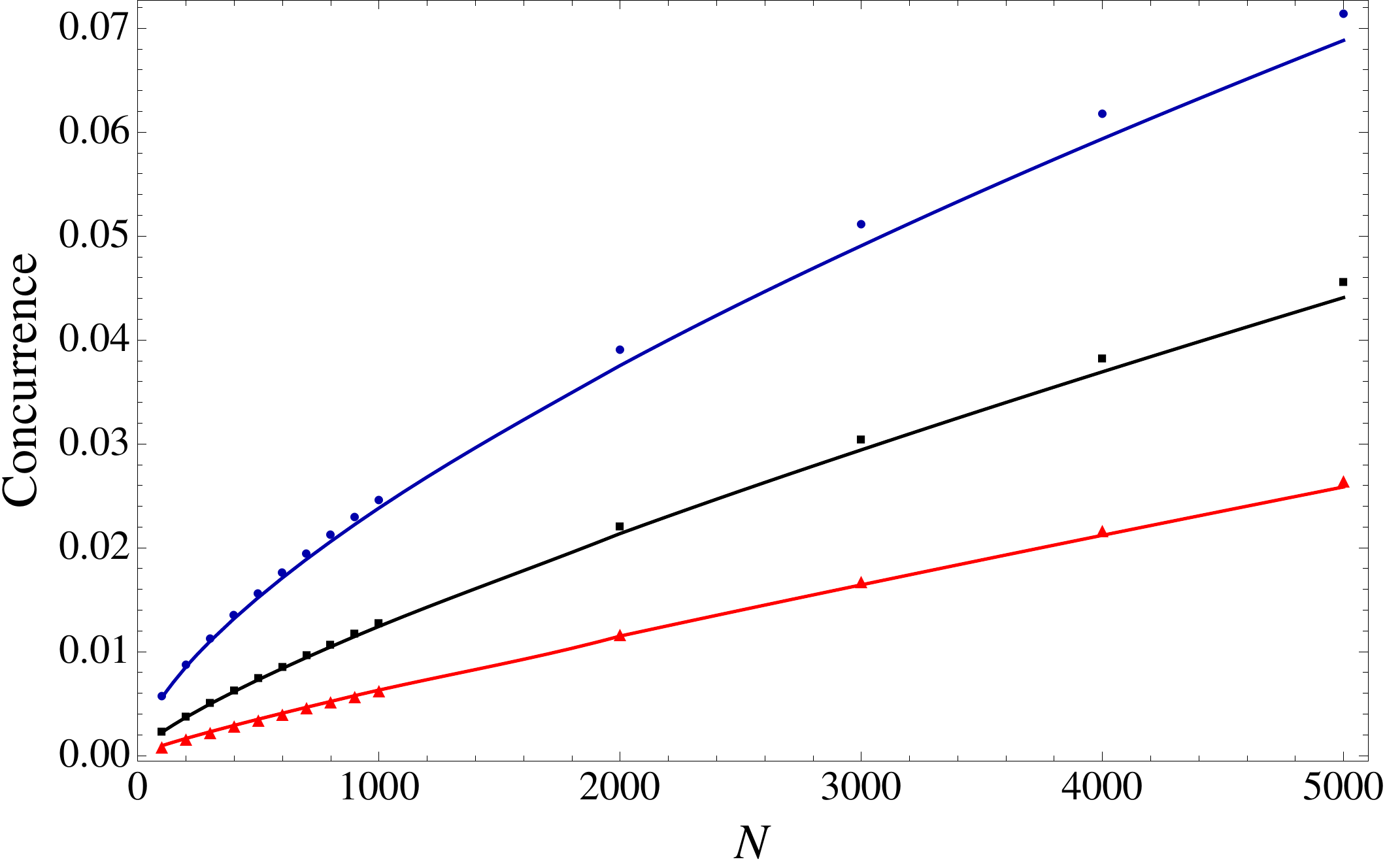}
\caption{(Color online) Concurrence of the condensate wave function as a function of the number of atoms in the condensate.  The discrete points are the concurrence calculated from the Schmidt coefficients found from the numerical solution of the 3D~GP equation, as described in the text: circles (blue) signify $q = 2$, squares (black) $q = 4$, and triangles (red) $q = 10$.  The corresponding lines are the concurrence of the perturbative Schmidt wave function, as given by Eq.~(\ref{concurrence}). The entanglement between transverse and longitudinal dimensions decreases for higher values of $q$, since the potential becomes more homogeneous.  Moreover, the entanglement remains remarkably small even for relatively large $N$, indicating that the condensate is well approximated by a single Schmidt~term.}
\label{fig:concurrence}
\end{figure}

In Fig.~\ref{fig:concurrence}, we compare the concurrence of the exact 3D solution, given by $\tilde{C}=2\tilde{c}_0\tilde{c}_1$, with the concurrence~(\ref{concurrence}) of the perturbative Schmidt theory~\cite{note:concurrence}.  One can see that the entanglement remains remarkably small even for relatively large atom numbers.  Notice that as the inhomogeneity of the longitudinal potential decreases, so does the spatial entanglement, which is an immediate consequence of Eq.~(\ref{phi10}).  In fact, for homogeneous longitudinal potentials, such as rings or boxes, the spatial entanglement vanishes completely.

The probability~(\ref{P}) and concurrence play complementary roles in assessing the accuracy of the perturbation theory: the deficit $P_D$ tells us to what extent the exact solution is confined to the two-dimensional subspace of the perturbation theory, and the concurrence $C$ tells us whether the perturbation theory captures correctly the entanglement in this subspace.

\begin{figure}
\includegraphics[scale=.4]{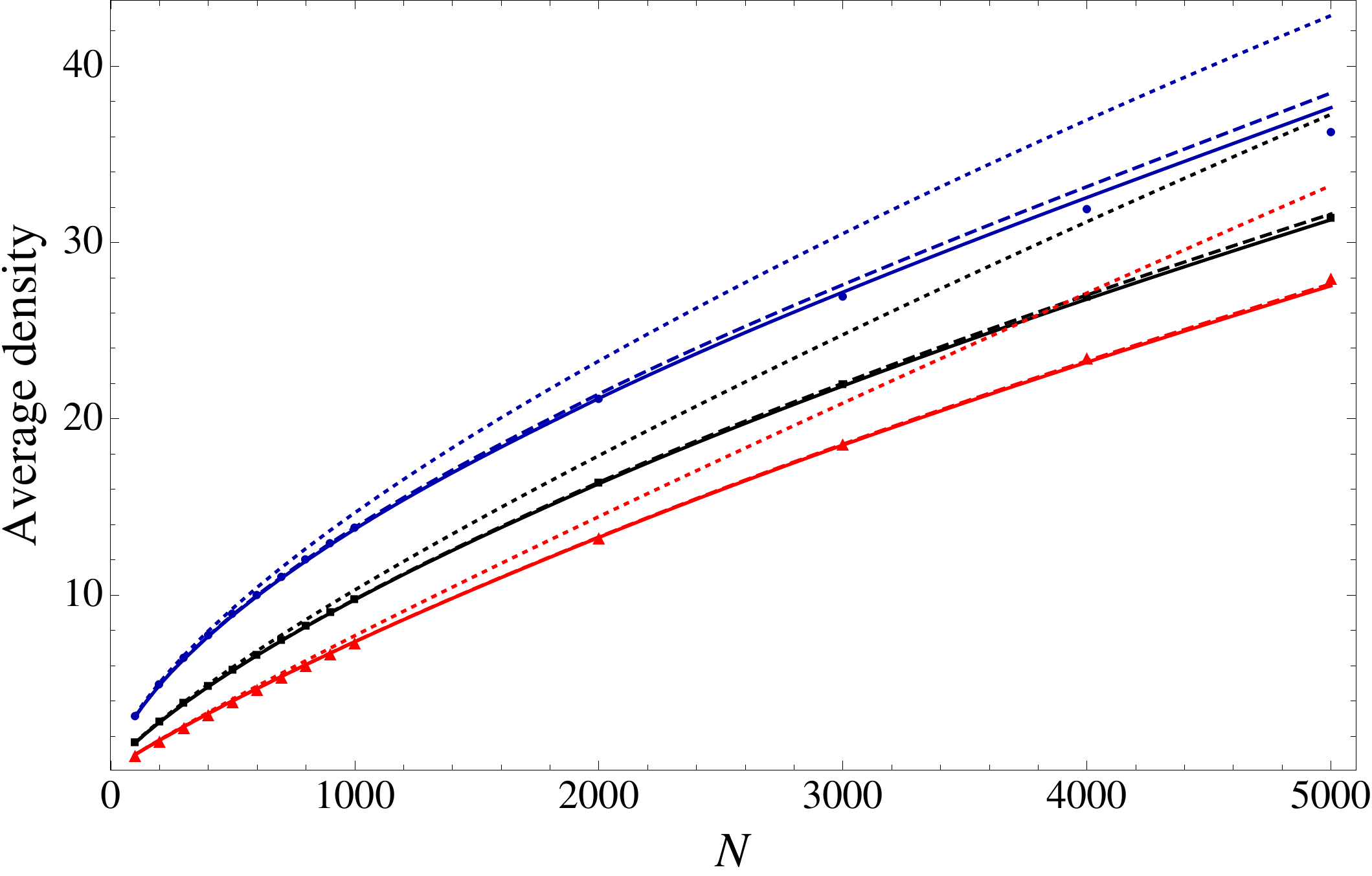}
\caption{(Color online) Average atomic density $N\eta$ in harmonic trap units as a function of the number of atoms in the condensate.  Discrete points are the results obtained from the numerical solution of the 3D~GP equation for different trap geometries: circles (blue) are for $q=2$, squares (black) for $q=4$, and triangles (red) for $q=10$.  The corresponding solid lines are those obtained from the entire perturbative Schmidt wave function~(\ref{psiSch}), and the dashed lines are those of the dominant Schmidt term only, whereas the dotted lines, which deviate substantially from the exact numerical results, are the result, $N\eta_T\eta_L$, obtained from the quasi-1D approximation.  The agreement between the exact results and those of the perturbation theory, even for relatively large atom numbers, is remarkable.  The difference between the solid and dashed lines comes from the nonseparable corrections to the dominant Schmidt term; this contribution is quite small, confirming the conclusions drawn from the concurrence.}
\label{fig:avgdensity}
\end{figure}

We can also use another, more directly physical quantity to quantify the effects of the nonseparable correction, namely, the condensate average density $N\eta = N\langle \psi | \psi^3\rangle$.  This parameter is of special interest in interferometric applications of BECs~\cite{tacla10}.  The average density is particularly appealing, because it can be used to measure any of the perturbative corrections introduced by the Schmidt decomposition, not just the nonseparable corrections.  In Fig.~\ref{fig:avgdensity}, we plot the average density obtained from the three-dimensional GP ground-state solution, as well as its estimates obtained from (i)~the entire perturbative Schmidt wave function~(\ref{psiSch}), (ii)~the dominant Schmidt term alone, $\chi_0\phi_0$, and (iii)~the quasi-1D, reduced-dimension approximation, which gives average density $N\eta_L\eta_T$.  Notice that the approximation provided by the Schmidt decomposition is remarkably good even for relatively large atom numbers.  The same is true for the approximation given by the dominant Schmidt term, which demonstrates how small is the effect of the nonseparable corrections.  Not surprisingly, in view of our previous results, the unperturbed estimate $N\eta_L\eta_T$ quickly deviates from the exact numerical results.

The results of this subsection point to the interesting conclusion that the nonseparable corrections to the dominant Schmidt term are small even for relatively large $N$. Therefore, as far as the ground-state properties go, one can nearly neglect the spatial entanglement and describe the condensate wave function in terms of the product wave function corresponding to the first Schmidt term.


\section{Conclusion}
\label{sec:conclusion}

We begin our concluding section by noting that there is another way to do the perturbation theory, which we call the relative-state method, in contrast to the Schmidt-decomposition method developed in this paper.  The relative-state method starts from a relative-state decomposition of the condensate wave function in the bare transverse eigenbasis $\xi_n$:
\begin{equation}
\psi(\bm{\rho},\bm{r})=\xi_0(\bm{\rho})\varphi_0(\bm{r})+
\epsilon\sum_{n=1}^\infty \xi_n(\bm{\rho})\varphi_n(\bm{r})\;.
\end{equation}
The dominant longitudinal wave function, $\varphi_0$, is determined by projecting the 3D~GP equation~(\ref{timeindepGPE}) onto $\xi_0$ and thus satisfies the reduced-dimension GP equation~(\ref{dDGPE}); it is not the same as $\phi_0$ of the Schmidt theory, but rather is the same as $\phi_{00}$.  The relative-state longitudinal wave functions $\varphi_n$ are neither orthogonal nor normalized.  One can work out the relative-state perturbation theory order by order, and this is more straightforward than the same process in the Schmidt-decomposition perturbation theory.  At the order we are working, we have shown explicitly that the two perturbation theories are equivalent.

Even though the relative-state perturbation theory is easier to implement, especially at higher orders than we consider in this paper, it obscures the physics of the condensate wave function.  In the Schmidt perturbation theory, the $n=0$ term gives the best product approximation to the exact wave function, and the Schmidt theory gets directly at how that term is modified by higher-order effects of the nonlinear scattering interaction.  The $n>0$ terms in the Schmidt theory describe entanglement of the transverse and longitudinal directions, and the Schmidt approach neatly separates this entanglement from the dominant ($n=0$) Schmidt term.  The relative-state theory can be used to calculate all these effects---and, as noted, it gives the same results at the order we are working---but it does not divide up the terms in perturbation theory in this neatly interpreted way.

We close with two ideas, on both of which we have already commenced work.  The comparisons of the Schmidt perturbation theory with the exact numerical results, presented in Sec.~\ref{sec:simulations}, suggest that the ground-state condensate wave function, for surprisingly high atom numbers, is well approximated by the two-Schmidt-term perturbation theory and even by the dominant Schmidt term alone.  This prompts us to wonder what happens to the Schmidt decomposition of the ground-state wave function as the atom number crosses over from reduced-dimension behavior for atom numbers below $N_T$ to the three-dimensional behavior above $N_T$.  We have evidence, to be presented elsewhere, that throughout the cross-over, the ground-state wave function continues to be well approximated by two Schmidt terms, and this suggests that an approximate description based on just two Schmidt terms might work through the entire cross-over and into the three-dimensional regime.

As we noted out above, the corrections to the reduced-dimension approximation are particularly important for interferometric schemes that use BECs~\cite{tacla10}.  A remaining problem in that regard is to assess how such corrections propagate in time and affect the condensate dynamics in a highly anisotropic trap.  We plan to address these questions in a future publication, which will develop the Schmidt perturbation theory for the time-dependent GP equation.


\acknowledgments
The authors thank Joshua Combes for help with plotting some of the results.
This work was supported in part by the National Science Foundation
(Grant Nos.~PHY-0903953 and~PHY-1005540).

\appendix

\vspace{6pt}
\section{Perturbation expansion}
\label{ap:expansion}

Our objective in this Appendix is to derive the equations for $\chi_{00}$, $\phi_{00}$, $\chi_{01}$, $\phi_{01}$, $\chi_{10}$, and $\phi_{10}$.  The first two of these give the lowest-order contribution to the dominant ($n=0$) Schmidt term, and the second two are the first-order corrections to the lowest-order behavior of this dominant Schmidt term.  The last two give the lowest-order contribution to the second Schmidt term and thus describe the lowest-order entanglement between the transverse and longitudinal directions.

We only need to work to at most second order in $\epsilon$ to determine the quantities we are interested in, so we use the Schmidt decomposition~(\ref{Schmidt}) in the form
\begin{equation}
\psi=\chi_0\phi_0+\epsilon\chi_1\phi_1+\epsilon^2\chi_2\phi_2+O(\epsilon^3)\;.
\label{Schmidt2}
\end{equation}
The transverse and longitudinal basis functions are further expanded as in Eqs.~(\ref{expandchin}) and~(\ref{expandphin}).

\begin{widetext}

We first consider the consequences of the various normalization and orthogonality conditions.  The overall normalization of the wave function,
\begin{align}
1=\langle\psi|\psi\rangle&=\langle\chi_0|\chi_0\rangle+\epsilon^2\langle\chi_1|\chi_1\rangle+O(\epsilon^4)\nonumber\\
&=\langle\chi_{00}|\chi_{00}\rangle+
2\epsilon\langle\chi_{01}|\chi_{00}\rangle+
\epsilon^2\bigl(2\langle\chi_{02}|\chi_{00}\rangle+\langle\chi_{01}|\chi_{01}\rangle+\langle\chi_{10}|\chi_{10}\rangle\bigr)
+O(\epsilon^4)\;,
\end{align}
implies that
\begin{align}
\langle\chi_{00}|\chi_{00}\rangle&=1\;,\\
\langle\chi_{01}|\chi_{00}\rangle&=0\;,\\
\langle\chi_{02}|\chi_{00}\rangle&=-{\langle\chi_{01}|\chi_{01}\rangle+\langle\chi_{10}|\chi_{10}\rangle\over2}\;.
\end{align}

The orthogonality relations~(\ref{chinchim}) for the transverse basis functions, carried to second order in $\epsilon$, imply
\begin{align}
c_n^2\delta_{nm}=\langle\chi_n|\chi_m\rangle=\langle\chi_{n0}|\chi_{m0}\rangle
+\epsilon\bigl(\langle\chi_{n1}|\chi_{m0}\rangle+\langle\chi_{n0}|\chi_{m1}\rangle\bigr)
+\epsilon^2\bigl(\langle\chi_{n2}|\chi_{m0}\rangle+\langle\chi_{n0}|\chi_{m2}\rangle+\langle\chi_{n1}|\chi_{m1}\rangle\bigr)
+O(\epsilon^3)\;.
\end{align}
For the cases of interest, we get
\begin{align}
n=m=0&:\quad c_0^2=\langle\chi_0|\chi_0\rangle=1-\epsilon^2\langle\chi_{10}|\chi_{10}\rangle+O(\epsilon^3)\;,\\
n=1,m=0&:\quad\langle\chi_{10}|\chi_{00}\rangle=0\;,\quad
\langle\chi_{11}|\chi_{00}\rangle=-\langle\chi_{10}|\chi_{01}\rangle\;,\\
n=2,m=0&:\quad\langle\chi_{20}|\chi_{00}\rangle=0\;,\\
n=m=1&:\quad c_1^2=\langle\chi_1|\chi_1\rangle=\langle\chi_{10}|\chi_{10}\rangle+O(\epsilon)\;.
\end{align}
Likewise, the orthonormality~(\ref{phinphim}) of the longitudinal basis functions, carried to the same order in $\epsilon$, gives
\begin{align}
n=m=0&:\quad\langle\phi_{00}|\phi_{00}\rangle=1\;,\quad
\langle\phi_{01}|\phi_{00}\rangle=0\;,\quad
\langle\phi_{02}|\phi_{00}\rangle=-\langle\phi_{01}|\phi_{01}\rangle/2\;,\\
n=1,m=0&:\quad\langle\phi_{10}|\phi_{00}\rangle=0\;,\quad
\langle\phi_{11}|\phi_{00}\rangle=-\langle\phi_{10}|\phi_{01}\rangle\;,\\
n=2,m=0&:\quad\langle\phi_{20}|\phi_{00}\rangle=0\;,\\
n=m=1&:\quad\langle\phi_{10}|\phi_{10}\rangle=1\;.
\end{align}

Now we use the Schmidt decomposition~(\ref{Schmidt2}) to expand the GP equation~(\ref{3DGPE}) to second order in powers of $\epsilon$, i.e.,
\begin{align}
\label{Aexpansion-3DGPE}
(\mu_0+&\epsilon\mu_1+\epsilon^2\mu_2)\chi_0\phi_0
+ (\mu_0 +\epsilon\mu_1)\epsilon \chi_1\phi_1
+ \mu_0\epsilon^2 \chi_2\phi_2\nonumber\\
&= H_T(\chi_0\phi_0 + \epsilon \chi_1\phi_1 + \epsilon^2 \chi_2\phi_2)
+ \epsilon( H_L + \tilde g\chi_0^2\phi_0^2) \chi_0\phi_0
+ \epsilon^2 ( H_L + 3 \tilde g\chi_0^2\phi_0^2 ) \chi_1\phi_1 + O(\epsilon^3)\;.
\end{align}
By projecting Eq.~(\ref{Aexpansion-3DGPE}) onto $\phi_0$ and then onto $\chi_0$, keeping in mind the strict orthogonality and reality of the Schmidt basis functions, we get
\begin{align}
(H_T &- \mu_0) \chi_0
+ \epsilon \big[\langle \phi_0 |(H_L - \mu_1)| \phi_0 \rangle + \tilde g \langle \phi_0 | \phi_0^3 \rangle \chi_0^2 \big]\chi_0\nonumber\\
&+ \epsilon^2 \big[ -\mu_2 \chi_0
+ \langle \phi_0 | (H_L - \mu_1) | \phi_1 \rangle \chi_1 + 3 \tilde g \langle \phi_1 | \phi_0^3 \rangle \chi_0^2 \chi_1 \big] + O(\epsilon^3) = 0\;,\\
\label{Aprojection-3DGPEchi0}
\langle \chi_0 |(H_T &- \mu_0)| \chi_0 \rangle \phi_0
+ \epsilon \big[ \langle \chi_0 |(H_T-\mu_0)| \chi_1 \rangle \phi_1
+ \langle\chi_0|\chi_0\rangle \big(H_L - \mu_1)\phi_0
+ \tilde g\langle \chi_0 | \chi_0^3 \rangle \phi_0^3 \big]\nonumber\\
&+ \epsilon^2 \big[ -\langle\chi_0|\chi_0\rangle \mu_2 \phi_0
+ \langle \chi_0 | (H_T - \mu_0) | \chi_2 \rangle \phi_2
+ 3 \tilde g\langle \chi_1|\chi_0^3 \rangle \phi_0^2 \phi_1 \big] + O(\epsilon^3) = 0\;.
\end{align}

We now expand the Schmidt-basis functions in powers of $\epsilon$ as in Eqs.~(\ref{expandchin}) and~(\ref{expandphin}).  Keeping terms to second order, we find that the transverse part of the first Schmidt term is determined by
\begin{align}
(\mu_0 - H_T) \chi_{00} &= 0\;, \label{Achi00-1} \\
(\mu_0 - H_T) \chi_{01} &= \bigl[\langle \phi_{00} |(H_L - \mu_1)| \phi_{00} \rangle
+ \tilde g \eta_{L}\chi_{00}^2\bigr]\chi_{00}\;,\label{Achi01-1}\\
(\mu_0 - H_T) \chi_{02} &=
\bigl[ 2\langle \phi_{01} |(H_L - \mu_1)| \phi_{00} \rangle
+ 4 \tilde g \langle \phi_{01} | \phi_{00}^3 \rangle \chi_{00}^2 - \mu_2 \bigr]\chi_{00}\nonumber\\
&\qquad + \big[ \langle \phi_{00} |(H_L-\mu_1)| \phi_{00} \rangle +
3 \tilde g \eta_{L} \chi_{00}^2 \big]\chi_{01}
+ \bigl[ \langle \phi_{10} | (H_L-\mu_1) | \phi_{00} \rangle +
3 \tilde g \langle \phi_{10} | \phi_{00}^3 \rangle\chi_{00}^2 \bigr]\chi_{10}\;,
	 \label{Achi02-1}
\end{align}
and the longitudinal part by
\begin{align}
\langle\chi_{00}|(H_T-\mu_0)|\chi_{00}\rangle\phi_{00}&=0\;,\\
\big[\langle\chi_{00}|\chi_{00}\rangle(\mu_1 - H_L) - \tilde g \eta_{T}\phi_{00}^2\big] \phi_{00}
&= 2 \langle \chi_{01} |(H_T-\mu_0)| \chi_{00} \rangle \phi_{00}
+ \langle \chi_{00} |(H_T-\mu_0)| \chi_{00} \rangle \phi_{01} \nonumber\\
&\qquad+ \langle \chi_{10} |(H_T-\mu_0)| \chi_{00} \rangle \phi_{10}\;,\\
\big[\langle\chi_{00}|\chi_{00}\rangle(\mu_1 - H_L) - 3 \tilde g \eta_{T} \phi_{00}^2 \big] \phi_{01}
&= \big[\langle \chi_{01} |(H_T-\mu_0)| \chi_{01} \rangle + 2 \langle \chi_{02} |(H_T-\mu_0)| \chi_{00} \rangle
+ 2\langle\chi_{01}|\chi_{00}\rangle(H_L-\mu_1)\nonumber\\
&\qquad\quad+ 4 \tilde g \langle \chi_{01} | \chi_{00}^3 \rangle \phi_{00}^2
- \langle\chi_{00}|\chi_{00}\rangle\mu_2 \big] \phi_{00} \nonumber\\
&\qquad + 2 \langle \chi_{01} |(H_T-\mu_0)| \chi_{00} \rangle \phi_{01} \nonumber\\
&\qquad + \big[ \langle \chi_{01} |(H_T-\mu_0)| \chi_{10} \rangle
+ \langle \chi_{11} |(H_T-\mu_0)| \chi_{00} \rangle
+ 3 \tilde g \langle \chi_{10} | \chi_{00}^3 \rangle \phi_{00}^2 \big] \phi_{10}\nonumber \\
&\qquad + \langle \chi_{00} |(H_T-\mu_0)| \chi_{00} \rangle \phi_{02}
+ \langle \chi_{10} |(H_T-\mu_0)| \chi_{00} \rangle \phi_{11}\nonumber \\
&\qquad+ \langle \chi_{20} |(H_T-\mu_0)| \chi_{00} \rangle \phi_{20}\;,
\end{align}
where $\eta_{T}$ and $\eta_{L}$, defined in Eqs.~(\ref{etaT}) and~(\ref{etaL}), are the lowest-order terms in the expansions of $\langle \chi_0 | \chi_0^3 \rangle$ and $\langle \phi_0 | \phi_0^3 \rangle$.   We now use the normalization and orthogonality conditions from above and the transverse equations to discard the first longitudinal equation and to simplify considerably the other two:
\begin{align}\label{Aphi00-1}
		(\mu_1 - H_L -\tilde g \eta_{T}\phi_{00}^2)\phi_{00}&=0\;,\\
		(\mu_1 - H_L - 3 \tilde g \eta_{T} \phi_{00}^2)\phi_{01}
&= \big[\langle \chi_{01} |(H_T-\mu_0)| \chi_{01} \rangle
+ 4\tilde g \langle \chi_{01} | \chi_{00}^3 \rangle \phi_{00}^2
-\mu_2 \big] \phi_{00} \nonumber\\
&\qquad + \big[ \langle \chi_{01} |(H_T-\mu_0)| \chi_{10} \rangle
+ 3 \tilde g \langle \chi_{10} | \chi_{00}^3 \rangle  \phi_{00}^2 \big] \phi_{10}\;.\label{Aphi01-1}
\end{align}

The lowest-order ($m=0$) equations for the dominant ($n=0$) Schmidt term are the transverse Schr\"odinger equation~(\ref{Achi00-1}) for $\chi_{00}$ and a longitudinal GP equation~(\ref{Aphi00-1}) for $\phi_{00}$. Both $\chi_{00}$ and $\phi_{00}$ are normalized to unity.  These lowest-order equations are precisely those that give the reduced-dimension approximation discussed in Sec.~\ref{sec:lowDBECs}.

We can now use the longitudinal equations to simplify the transverse equations and these simplified transverse equations, in turn, to simplify further the longitudinal equations, the results being
\begin{align}
	(\mu_0 - H_T) \chi_{00} &= 0\;, \label{Achi00} \\
	(\mu_0 - H_T) \chi_{01} &= \tilde g\eta_L(\chi_{00}^2-\eta_T)\chi_{00}\;,\label{Achi01}\\
	(\mu_0 - H_T) \chi_{02} &=  \bigl[ 2\tilde g\langle\phi_{01}|\phi_{00}^3\rangle(2\chi_{00}^2-\eta_T)-\mu_2\bigr]\chi_{00}
+ \tilde g\eta_L(3\chi_{00}^2-\eta_T)\chi_{01}
+ \tilde g\langle\phi_{10}|\phi_{00}^3\rangle(3\chi_{00}^2-\eta_T)\chi_{10}\;,
\label{Achi02}\\
    \label{Aphi00}
	(\mu_1 - H_L -\tilde g \eta_{T}\phi_{00}^2)\phi_{00}&=0\;,\\
	(\mu_1 - H_L - 3 \tilde g \eta_{T} \phi_{00}^2)\phi_{01}
&= \big[\tilde g \langle \chi_{01} | \chi_{00}^3 \rangle (4 \phi_{00}^2-\eta_L)-\mu_2 \big] \phi_{00}
+ \tilde g \langle \chi_{10} | \chi_{00}^3 \rangle ( 3 \phi_{00}^2 -\eta_L) \phi_{10}\;.\label{Aphi01-02}
\end{align}
Notice that the right-hand side of Eq.~(\ref{Achi01}) is orthogonal to $\chi_{00}$, as required by the left-hand side.

The remaining first-order terms, $\chi_{10}$ and $\phi_{10}$, provide the first correction to a product wave function and thus describe to lowest order the entanglement of the transverse and longitudinal dimensions.  These terms are determined by projecting Eq.~(\ref{Aexpansion-3DGPE}) onto $\phi_1$ and $\chi_1$.  The first of these gives
\begin{equation}
(H_T - \mu_0)\chi_1 + \langle \phi_1 |(H_L - \mu_1)| \phi_0 \rangle \chi_0
+ \tilde g \langle \phi_1 |\phi_0^3 \rangle \chi_0^3 + O(\epsilon) = 0\;.
\end{equation}
Plugging in the expansions~(\ref{expandchin}) and~(\ref{expandphin}), we find that
\begin{equation}
\label{Achi10}
	(\mu_0 - H_T) \chi_{10}
= \bigl[ \langle \phi_{10} |(H_L - \mu_1)| \phi_{00} \rangle
+ \tilde g \langle \phi_{10} |\phi_{00}^3 \rangle \chi_{00}^2 \bigr]\chi_{00}
= \tilde g \langle \phi_{10} |\phi_{00}^3 \rangle ( \chi_0^2 - \eta_{T}) \chi_{00} \;,
\end{equation}
where to obtain the second form, we use the longitudinal equation~(\ref{Aphi00}) for $\phi_{00}$.  The right-hand side of Eq.~(\ref{Achi10}) is orthogonal to $\chi_{00}$, as required by the left-hand side.

Projecting Eq.~(\ref{Aexpansion-3DGPE}) onto $\chi_1$ yields
\begin{eqnarray}
	\langle \chi_1 |(H_T-\mu_0)| \chi_0\rangle \phi_0
+ \epsilon \big[ \langle \chi_1 |(H_T - \mu_0) |\chi_1\rangle \phi_1+
\tilde g \langle \chi_1 |\chi_0^3 \rangle \phi_0^3 \big] + O(\epsilon^2) = 0\;.
\end{eqnarray}
Applying the expansions~(\ref{expandchin}) and~(\ref{expandphin}) gives at lowest order, $\langle\chi_{10}|(H_T-\mu_0)|\chi_{00}\rangle\phi_{00}=0$, which is already satisfied, and at the next order,
\begin{align}
	\langle \chi_{10} |(\mu_0-H_T)| \chi_{10}\rangle \phi_{10}
&=\bigl[\langle\chi_{11}|(H_T-\mu_0)|\chi_{00}\rangle
+\langle\chi_{10}|(H_T-\mu_0)|\chi_{01}\rangle
+ \tilde g \langle \chi_{10} |\chi_{00}^3 \rangle \phi_{00}^2\bigr]\phi_{00}\nonumber\\
&\qquad + \langle \chi_{10} |(H_T - \mu_0) |\chi_{00}\rangle \phi_{01}\;.
\end{align}

\end{widetext}

\noindent
Simplifying this equation using Eqs.~(\ref{Achi00}), (\ref{Achi01}), and (\ref{Achi10}) gives a remarkably simple algebraic equation for $\phi_{10}$,
\begin{equation}
\langle\phi_{10}|\phi_{00}^3\rangle\phi_{10}=\phi_{00}^3-\eta_L\phi_{00}\;.
\end{equation}
This means that $\phi_{00}^3$ is a linear combination of the orthonormal functions $\phi_{00}$ and $\phi_{10}$ and hence that
\begin{equation}
\langle\phi_{00}^3|\phi_{00}^3\rangle=\eta_L^2+\langle\phi_{10}|\phi_{00}^3\rangle^2\;.
\end{equation}
Since $\eta_L=\langle\phi_{00}|\phi_{00}^3\rangle$ is the average value of $\phi_{00}^2$ over the distribution $\phi_{00}^2$ and $\langle\phi_{00}^3|\phi_{00}^3\rangle$ is the second moment of $\phi_{00}^2$, $\langle\phi_{10}|\phi_{00}^3\rangle^2$ is the (nonegative) variance of $\phi_{00}^2$, which we denote as $\Delta\eta_L^2$.  Putting all this together, we have
\begin{align}\label{ADeltaeta}
&\langle\phi_{10}|\phi_{00}^3\rangle=\Delta\eta_L\equiv
\sqrt{\langle\phi_{00}^3|\phi_{00}^3\rangle-\eta_L^2}\;,\\
&\phi_{10}=\frac{\phi_{00}^2-\eta_L}{\Delta\eta_L}\phi_{00}\;.
\label{Aphi10}
\end{align}

We now let $\{\xi_n\}$ be the set of energy eigenfunctions of the bare transverse potential, with $E_n$ denoting the corresponding energy eigenvalues, i.e., $H_T\xi_n=E_n\xi_n$.  From Eq.~(\ref{Achi00}), we have $\chi_{00}=\xi_0$ and $\mu_0=E_0$.  Equations~(\ref{Achi01}) and~(\ref{Achi10}) imply that
\begin{equation}
\frac{\chi_{01}}{\eta_L}=\frac{\chi_{10}}{\Delta\eta_L}
=-\tilde g\sum_{n=1}^\infty \xi_n\frac{\langle\xi_n|\xi_0^3\rangle}{E_n-E_0}\;,
\end{equation}
and this, in turn, gives us the relations~(\ref{chi01chi003}).

The one remaining step we need to take is to use the relations~(\ref{chi01chi003}) and our result~(\ref{Aphi10}) for $\phi_{10}$ to simplify the equation, Eq.~(\ref{Aphi01-02}), for $\phi_{01}$:
\begin{equation}\label{Aphi01}
(\mu_1-H_L-3\tilde g\eta_T\phi_{00}^2)\phi_{01}
=3\tilde g^2\UpsilonT\phi_{00}^5-\mu_2\phi_{00}\;,
\end{equation}
where $\UpsilonT$ is the transverse coupling parameter defined by Eq.~(\ref{UpsilonT}).
Projecting this equation onto $\phi_{00}$---or, equivalently, projecting the equation for $\chi_{02}$, Eq.~(\ref{Achi02}), onto $\chi_{00}$---gives the expression~(\ref{mu2}) for the chemical potential.  Plugging this expression for $\mu_2$ back into the equation for $\phi_{01}$, Eq.~(\ref{Aphi01}), shows that it is a linear integro-differential equation for $\phi_{01}$.  Solving this integro-differential equation and using the result in Eq.~(\ref{mu2}) determines $\mu_2$.

This completes the set of equations we need.  Section~\ref{sec:decomposition} summarizes the final forms of our perturbative equations and also derives a GP-like, but quintic equation for $\phi_0=\phi_{00}+\epsilon\phi_{01}$, which we use in preference to the separate equations for $\phi_{00}$ and~$\phi_{01}$.


\begin{thebibliography}{20}
\bibitem{NT}
Instead of solving Eq.~(\ref{NT}) numerically, we use the product ansatz~(\ref{ansatz}) to calculate $N_T$ using the Thomas-Fermi approximation for the longitudinal wave function. The details of this calculation can be found in Ref.~\cite{tacla10}.
\bibitem{numerics}
We numerically integrate Eqs.~(\ref{timeindepGPE}) and~(\ref{phi00}) using the method described in Ref.~\cite{dion07}, whereas $\phi_0$ is found by imaginary time propagation.
\bibitem{note:concurrence}
In the definition of the concurrence, the states are assumed to be normalized, which would require the perturbative Schmidt coefficients, $c_0$ and $c_1$, and the coefficients $\tilde c_0$ and $\tilde c_1$ to be normalized to unity.  The perturbative Schmidt coefficients are already normalized at the order of perturbation theory we are working, so renormalizing them would only introduce meaningless higher-order corrections.  The data in Table~\ref{tab:probability} show that $\tilde c_0$ and $\tilde c_1$ are very close to normalized; renormalizing them has a negligible effect on the calculated concurrences.
\end{thebibliography}
\end{document}